\documentclass[twocolumn,twoside, superscriptaddress, slac_two]{revtex4}
\usepackage{caption}
\usepackage{graphicx}
\usepackage{fancyhdr}
\begin{document}

\title{The KASCADE-Grande experiment: measurements of the all-particle energy
  spectrum of cosmic rays}

\author{J.C. Arteaga-Vel\'azquez}
\email[corresponding author, email: ]{arteaga@ifm.umich.mx}
\altaffiliation[now at: ]{Universidad Michoacana, Instituto de F\'\i sica y Matem\'aticas, Mexico}
\affiliation{\leftskip0.5in \looseness-2 Institut f\"ur Experimentelle Kernphysik, Karlsruher Institut f\"ur Technologie - Campus S\"ud, Germany}
\author{W.D.~Apel}
\affiliation{\leftskip0.5in \looseness-2 Institut f\"ur Kernphysik, Karlsruher Institut f\"ur Technologie - Campus Nord, Germany}
\author{K.~Bekk}
\affiliation{\leftskip0.5in \looseness-2 Institut f\"ur Kernphysik, Karlsruher Institut f\"ur Technologie - Campus Nord, Germany}
\author{M.~Bertaina}
\affiliation{\leftskip0.5in \looseness-2 Dipartimento di Fisica Generale dell' Universita Torino, Italy}
\author{J.~Bl\"umer}
\affiliation{\leftskip0.5in \looseness-2 Institut f\"ur Kernphysik, Karlsruher Institut f\"ur Technologie - Campus Nord, Germany}
\affiliation{\leftskip0.5in \looseness-2 Institut f\"ur Experimentelle Kernphysik, Karlsruher Institut f\"ur Technologie - Campus S\"ud, Germany}
\author{H.~Bozdog}
\affiliation{\leftskip0.5in \looseness-2 Institut f\"ur Kernphysik, Karlsruher Institut f\"ur Technologie - Campus Nord, Germany}
\author{I.M.~Brancus}
\affiliation{\leftskip0.5in \looseness-2 National Institute of Physics and Nuclear Engineering, Bucharest, Romania}
\author{P.~Buchholz}
\affiliation{\leftskip0.5in \looseness-2 Fachbereich Physik, Universit\"at Siegen, Germany}
\author{E.~Cantoni}
\affiliation{\leftskip0.5in \looseness-2 Dipartimento di Fisica Generale dell' Universita Torino, Italy}
\affiliation{\leftskip0.5in \looseness-2 Istituto di Fisica dello Spazio Interplanetario, INAF Torino, Italy}
\author{A.~Chiavassa}
\affiliation{\leftskip0.5in \looseness-2 Dipartimento di Fisica Generale dell' Universita Torino, Italy}
\author{F.~Cossavella}
\altaffiliation[now at: ]{Max-Planck-Institut f\"ur Physik, M\"unchen, Germany}
\affiliation{\leftskip0.5in \looseness-2 Institut f\"ur Experimentelle Kernphysik, Karlsruher Institut f\"ur Technologie - Campus S\"ud, Germany}
\author{K.~Daumiller}
\affiliation{\leftskip0.5in \looseness-2 Institut f\"ur Kernphysik, Karlsruher Institut f\"ur Technologie - Campus Nord, Germany}
\author{V.~de Souza}
\altaffiliation[now at: ]{Universidade S$\tilde{a}$o Paulo, Instituto de F\'{\i}sica de S\~ao Carlos, Brasil}
\affiliation{\leftskip0.5in \looseness-2 Institut f\"ur Experimentelle Kernphysik, Karlsruher Institut f\"ur Technologie - Campus S\"ud, Germany}
\author{F.~Di~Pierro}
\affiliation{\leftskip0.5in \looseness-2 Dipartimento di Fisica Generale dell' Universita Torino, Italy}
\author{P.~Doll}
\affiliation{\leftskip0.5in \looseness-2 Institut f\"ur Kernphysik, Karlsruher Institut f\"ur Technologie - Campus Nord, Germany}
\author{R.~Engel}
\affiliation{\leftskip0.5in \looseness-2 Institut f\"ur Kernphysik, Karlsruher Institut f\"ur Technologie - Campus Nord, Germany}
\author{J.~Engler}
\affiliation{\leftskip0.5in \looseness-2 Institut f\"ur Kernphysik, Karlsruher Institut f\"ur Technologie - Campus Nord, Germany}
\author{M. Finger}
\affiliation{\leftskip0.5in \looseness-2 Institut f\"ur Kernphysik, Karlsruher Institut f\"ur Technologie - Campus Nord, Germany}
\author{D.~Fuhrmann}
\affiliation{\leftskip0.5in \looseness-2 Fachbereich Physik, Universit\"at Wuppertal, Germany}
\author{P.L.~Ghia}
\affiliation{\leftskip0.5in \looseness-2 Istituto di Fisica dello Spazio Interplanetario, INAF Torino, Italy}
\author{H.J.~Gils}
\affiliation{\leftskip0.5in \looseness-2 Institut f\"ur Kernphysik, Karlsruher Institut f\"ur Technologie - Campus Nord, Germany}
\author{R.~Glasstetter}
\affiliation{\leftskip0.5in \looseness-2 Fachbereich Physik, Universit\"at Wuppertal, Germany}
\author{C.~Grupen}
\affiliation{\leftskip0.5in \looseness-2 Fachbereich Physik, Universit\"at Siegen, Germany}
\author{A.~Haungs}
\affiliation{\leftskip0.5in \looseness-2 Institut f\"ur Kernphysik, Karlsruher Institut f\"ur Technologie - Campus Nord, Germany}
\author{D.~Heck}
\affiliation{\leftskip0.5in \looseness-2 Institut f\"ur Kernphysik, Karlsruher Institut f\"ur Technologie - Campus Nord, Germany}
\author{J.R.~H\"orandel}
\altaffiliation[now at: ]{Dept. of Astrophysics, Radboud University Nijmegen, The Netherlands}
\affiliation{\leftskip0.5in \looseness-2 Institut f\"ur Experimentelle Kernphysik, Karlsruher Institut f\"ur Technologie - Campus S\"ud, Germany}
\author{T.~Huege}
\affiliation{\leftskip0.5in \looseness-2 Institut f\"ur Kernphysik, Karlsruher Institut f\"ur Technologie - Campus Nord, Germany}
\author{P.G.~Isar}
\affiliation{\leftskip0.5in \looseness-2 Institut f\"ur Kernphysik, Karlsruher Institut f\"ur Technologie - Campus Nord, Germany}
\author{K.-H.~Kampert}
\affiliation{\leftskip0.5in \looseness-2 Fachbereich Physik, Universit\"at Wuppertal, Germany}
\author{D.~Kang}
\affiliation{\leftskip0.5in \looseness-2 Institut f\"ur Experimentelle Kernphysik, Karlsruher Institut f\"ur Technologie - Campus S\"ud, Germany}
\author{D.~Kickelbick}
\affiliation{\leftskip0.5in \looseness-2 Fachbereich Physik, Universit\"at Siegen, Germany}
\author{H.O.~Klages}
\affiliation{\leftskip0.5in \looseness-2 Institut f\"ur Kernphysik, Karlsruher Institut f\"ur Technologie - Campus Nord, Germany}
\author{K.~Link}
\affiliation{\leftskip0.5in \looseness-2 Institut f\"ur Experimentelle Kernphysik, Karlsruher Institut f\"ur Technologie - Campus S\"ud, Germany}
\author{P.~{\L}uczak}
\affiliation{\leftskip0.5in \looseness-2 Soltan Institute for Nuclear Studies, Lodz, Poland}
\author{M.~Ludwig}
\affiliation{\leftskip0.5in \looseness-2 Institut f\"ur Experimentelle Kernphysik, Karlsruher Institut f\"ur Technologie - Campus S\"ud, Germany}
\author{H.J.~Mathes}
\affiliation{\leftskip0.5in \looseness-2 Institut f\"ur Kernphysik, Karlsruher Institut f\"ur Technologie - Campus Nord, Germany}
\author{H.J.~Mayer}
\affiliation{\leftskip0.5in \looseness-2 Institut f\"ur Kernphysik, Karlsruher Institut f\"ur Technologie - Campus Nord, Germany}
\author{M.~Melissas}
\affiliation{\leftskip0.5in \looseness-2 Institut f\"ur Experimentelle Kernphysik, Karlsruher Institut f\"ur Technologie - Campus S\"ud, Germany}
\author{J.~Milke}
\affiliation{\leftskip0.5in \looseness-2 Institut f\"ur Kernphysik, Karlsruher Institut f\"ur Technologie - Campus Nord, Germany}
\author{B.~Mitrica}
\affiliation{\leftskip0.5in \looseness-2 National Institute of Physics and Nuclear Engineering, Bucharest, Romania}
\author{C.~Morello}
\affiliation{\leftskip0.5in \looseness-2 Istituto di Fisica dello Spazio Interplanetario, INAF Torino, Italy}
\author{G.~Navarra}
\altaffiliation[deceased]{}
\affiliation{\leftskip0.5in \looseness-2 Dipartimento di Fisica Generale dell' Universita Torino, Italy}
\author{S.~Nehls}
\affiliation{\leftskip0.5in \looseness-2 Institut f\"ur Kernphysik, Karlsruher Institut f\"ur Technologie - Campus Nord, Germany}
\author{J.~Oehlschl\"ager}
\affiliation{\leftskip0.5in \looseness-2 Institut f\"ur Kernphysik, Karlsruher Institut f\"ur Technologie - Campus Nord, Germany}
\author{S.~Ostapchenko}
\altaffiliation[now at: ]{University of Trondheim, Norway}
\affiliation{\leftskip0.5in \looseness-2 Institut f\"ur Kernphysik, Karlsruher Institut f\"ur Technologie - Campus Nord, Germany}
\author{S.~Over}
\affiliation{\leftskip0.5in \looseness-2 Fachbereich Physik, Universit\"at Siegen, Germany}
\author{N.~Palmieri}
\affiliation{\leftskip0.5in \looseness-2 Institut f\"ur Experimentelle Kernphysik, Karlsruher Institut f\"ur Technologie - Campus S\"ud, Germany}
\author{M.~Petcu}
\affiliation{\leftskip0.5in \looseness-2 National Institute of Physics and Nuclear Engineering, Bucharest, Romania}
\author{T.~Pierog}
\affiliation{\leftskip0.5in \looseness-2 Institut f\"ur Kernphysik, Karlsruher Institut f\"ur Technologie - Campus Nord, Germany}
\author{H.~Rebel}
\affiliation{\leftskip0.5in \looseness-2 Institut f\"ur Kernphysik, Karlsruher Institut f\"ur Technologie - Campus Nord, Germany}
\author{M.~Roth}
\affiliation{\leftskip0.5in \looseness-2 Institut f\"ur Kernphysik, Karlsruher Institut f\"ur Technologie - Campus Nord, Germany}
\author{H.~Schieler}
\affiliation{\leftskip0.5in \looseness-2 Institut f\"ur Kernphysik, Karlsruher Institut f\"ur Technologie - Campus Nord, Germany}
\author{F.~Schr\"oder}
\affiliation{\leftskip0.5in \looseness-2 Institut f\"ur Kernphysik, Karlsruher Institut f\"ur Technologie - Campus Nord, Germany}
\author{O.~Sima}
\affiliation{\leftskip0.5in \looseness-3 Department of Physics, University of Bucharest, Bucharest, Romania}
\author{G.~Toma}
\affiliation{\leftskip0.5in \looseness-2 National Institute of Physics and Nuclear Engineering, Bucharest, Romania}
\author{G.C.~Trinchero}
\affiliation{\leftskip0.5in \looseness-2 Istituto di Fisica dello Spazio Interplanetario, INAF Torino, Italy}
\author{H.~Ulrich}
\affiliation{\leftskip0.5in \looseness-2 Institut f\"ur Kernphysik, Karlsruher Institut f\"ur Technologie - Campus Nord, Germany}
\author{A.~Weindl}
\affiliation{\leftskip0.5in \looseness-2 Institut f\"ur Kernphysik, Karlsruher Institut f\"ur Technologie - Campus Nord, Germany}
\author{J.~Wochele}
\affiliation{\leftskip0.5in \looseness-2 Institut f\"ur Kernphysik, Karlsruher Institut f\"ur Technologie - Campus Nord, Germany}
\author{M.~Wommer}
\affiliation{\leftskip0.5in \looseness-2 Institut f\"ur Kernphysik, Karlsruher Institut f\"ur Technologie - Campus Nord, Germany}
\author{J.~Zabierowski}
\affiliation{\leftskip0.5in \looseness-2 Soltan Institute for Nuclear Studies, Lodz, Poland}

\begin{abstract}
The all-particle energy spectrum as measured by the KASCADE-Grande
 experiment for $E=10^{16} - 10^{18} \, \mbox{eV}$ is presented within
 the framework of the QGSJET II/FLUKA hadronic interaction models. Three
 different methods were applied based on the  muon size and the total 
 number of charged particles individually and in combination. From the
 study it is found that the spectrum cannot be completely
 described by a smooth power law due to the presence of characteristic
 features.
\end{abstract}

\maketitle
\thispagestyle{fancy}

 \section{INTRODUCTION}

 Powerful cosmic ray accelerators are hidden inside our own galaxy and in the heart
 of extragalactic objects, but their precise location and exact working mechanism 
 are still subject of intense debate. Cosmic ray research provides important clues through 
 the measurement of the energy spectrum, the arrival direction and the composition of
 such enigmatic particles by means of the extensive air showers (EAS) that cosmic rays
 produce in the Earth atmosphere. Along the past years, dedicated experiments have
 shown that the energy spectrum of cosmic rays extends from a few MeV up to
 $10^{20} \, \mbox{eV}$, has a striking power-law behavior with a spectral index $\gamma \approx -2.7$ 
 and exhibits some intriguing structures around $3-5 \cdot 10^{15}$ and $4-10 \cdot 10^{18}$
 eV, better known as the {\it knee} and the {\it ankle}, where the spectral index of the particle 
 flux decreases and grows, respectively \cite{Nagano}. The precise interpretation of these features are 
 not yet clear, but they could arise as a consequence of an interplay among different factors
 such as the loss of efficiency of the galactic accelerators, the dominance of a new
 component of extragalactic origin at the high-energy regime and the propagation 
 effects of cosmic rays through space \cite{Hillas, Bere}. To test different scenarios, accurate and 
 high-statistics measurements are needed between $10^{15}$ and $10^{18}$ eV. Although 
 observations of this kind have been already performed around the knee, the picture is not 
 yet complete due to the lack of quality EAS data at higher energies: $E = 10^{16} - 10^{18} 
 \, \mbox{eV}$. The goal of the KASCADE-Grande experiment is to close this gap between 
 the \textit{knee} and the ultrahigh-energy region with accurate EAS measurements.

 KASCADE-Grande is a ground-based EAS array integrated by different detection systems to measure
 and separate the muon and electromagnetic components of the EAS
 \cite{KG}. Once combined, both observables  become a powerful tool to
 reconstruct the energy spectrum and to estimate the composition of 
 primary cosmic ray events. In the following lines, the energy spectrum of cosmic rays for
 $E = 10^{16} - 10^{18} \, \mbox{eV}$ as measured by KASCADE-Grande will be 
 presented and the reconstruction method will be described. First, a brief introduction to the 
 characteristics and performance of the experiment will be given.

 \section{The KASCADE-Grande experiment}
  \subsection{Description}

  \begin{table}[t]
  \begin{center}
  \captionsetup{justification=RaggedRight}
  \caption{\small Some components of KASCADE-Grande: total sensitive areas and threshold 
   kinematic energies for vertically incident particles are presented. MTD refers to the muon
   tracking detectors \cite{KG}.}
  \begin{tabular}{l c c c }
  \hline \textbf{\footnotesize Detector} & \textbf{\footnotesize Particle} & \textbf{\footnotesize Area(m${}^2$)} &
  \textbf{\footnotesize Threshold}
  \\
\hline {\footnotesize Grande array} &&&\\
       {\footnotesize (plastic scintillators)}  & {\footnotesize Charged}    & {\footnotesize $370$} & {\footnotesize $3$ MeV} \\
       {\footnotesize Piccolo array}&&&\\
       {\footnotesize (plastic scintillators)}  & {\footnotesize Charged}    & {\footnotesize $80$}  & {\footnotesize $3$ MeV} \\
       {\footnotesize KASCADE array}&&&\\
       {\footnotesize (liquid scintillators)}   & {\footnotesize e$/\gamma$} & {\footnotesize $490$} & {\footnotesize $5$ MeV} \\
       {\footnotesize KASCADE array}&&&\\ 
       {\footnotesize (shielded plast. scint.)} & {\footnotesize $\mu$}      & {\footnotesize $622$} & {\footnotesize $230$ MeV} \\ 
       {\footnotesize MTD (streamer tubes)}     & {\footnotesize $\mu$}      & {\footnotesize $4 \times 128$} & {\footnotesize $800$ MeV} \\
\hline
 \end{tabular}
 \label{Tab1}
 \end{center}
\end{table}

\begin{figure}[!b]
 \begin{center}
 \includegraphics[width=70mm, height=75mm]{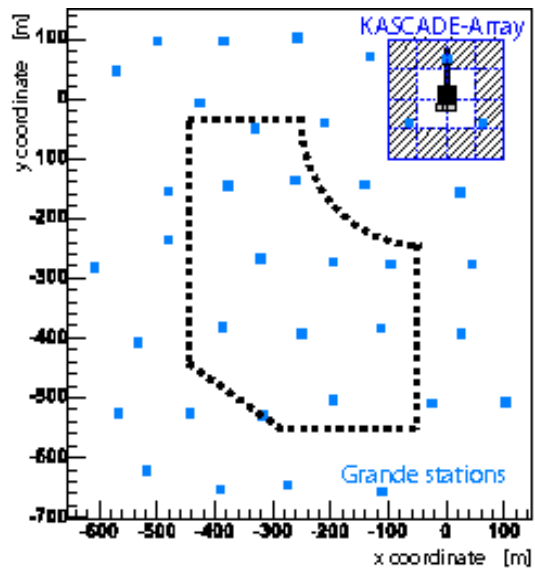}
 \captionsetup{justification=RaggedRight}
 \caption{\small The KASCADE-Grande experiment. Small squares represent the Grande stations. 
 The Kascade array is seen at the upper right hand of the figure. KASCADE detectors are 
 arranged in 16 clusters. The outer 12 clusters contain the muon detectors. The
 dotted line encloses the fiducial area selected for the present analysis.}
 \label{fig1}
 \end{center}
 \end{figure}

  The KASCADE-grande detector (located at $49.1^\circ$ N, $8.4^\circ$ E, 110 m a.s.l.) 
 is the successor of the KASCADE experiment \cite{KASCADE} and incorporates the original 
 electromagnetic detectors and muon devices of KASCADE to a bigger system 
 of detectors, called Grande, composed of a $700 \times 700 \, \mbox{m}^2$ array
 with $37 \times 10 \, \mbox{m}^2$ scintillator stations regularly spaced by an average 
 distance of $137 \, \mbox{m}$ \cite{KG}. To coordinate the trigger of KASCADE with EAS events 
 where the core is found inside Grande a smaller array, named Piccolo, was added. The 
 main characteristics of the KASCADE-Grande detectors are displayed in table \ref{Tab1}
 and the layout of the experiment is presented in figure \ref{fig1}.

 The Grande array is used to sample the density of charged particles 
 of the shower front at ground level and  to measure the particle arrival times
 of the EAS. The core position, the number of charged particles ($N_{ch}$) 
 and the arrival direction of the shower are extracted from the Grande data 
 through an iterative fit and a careful modeling of the EAS front. To reconstruct the 
 arrival direction a $\chi^2$ fit is applied assuming a curved shower 
 front as suggested by CORSIKA/QGSJET II simulations. On the other hand, to obtain the 
 core position and the shower size a maximum likelihood procedure was employed by 
 using a modified NKG lateral distribution function as a reference model \cite{KG}. 

 An important component of this experiment is the KASCADE muon array, composed by  
 $192 \times 3.2 \, \mbox{m}^2$ shielded scintillator detectors, which are sensitive 
 to muons with threshold energy above $230$ MeV for vertical incidence. 
 With the Grande information and the measurements of the shielded array of the lateral 
 distribution of muons in the shower front, the muon size ($N_\mu$) is reconstructed 
 event-by-event at KASCADE-Grande. The procedure involves the maximum loglikelihood 
 technique along with a Lagutin-Raikin distribution function \cite{KG}.

 \subsection{Accuracy of EAS Reconstruction}
 
 Systematic uncertainties for the core, $N_{ch}$ and arrival direction of the 
 EAS were studied directly \cite{KG} by comparing the results of the Grande and KASCADE 
 reconstructions, which work independently, for a subset of data with cores 
 located inside a common area and shower sizes in the interval $\log_{10}N_{ch} = 5.8-7.2$. 
 EAS core positions and arrival directions were
 found to be estimated by Grande with an accuracy of the order of $6 \, \mbox{m}$ and
 $0.8^{\circ}$, respectively. The same evaluation showed that the systematic uncertainties 
 of $N_{ch}$ at Grande are $\leq 5 \%$. Those values are in full agreement
 with expectations from Monte Carlo simulations.

 \begin{figure*}[!t]
 \centering
 \includegraphics[width=75mm, height=60mm]{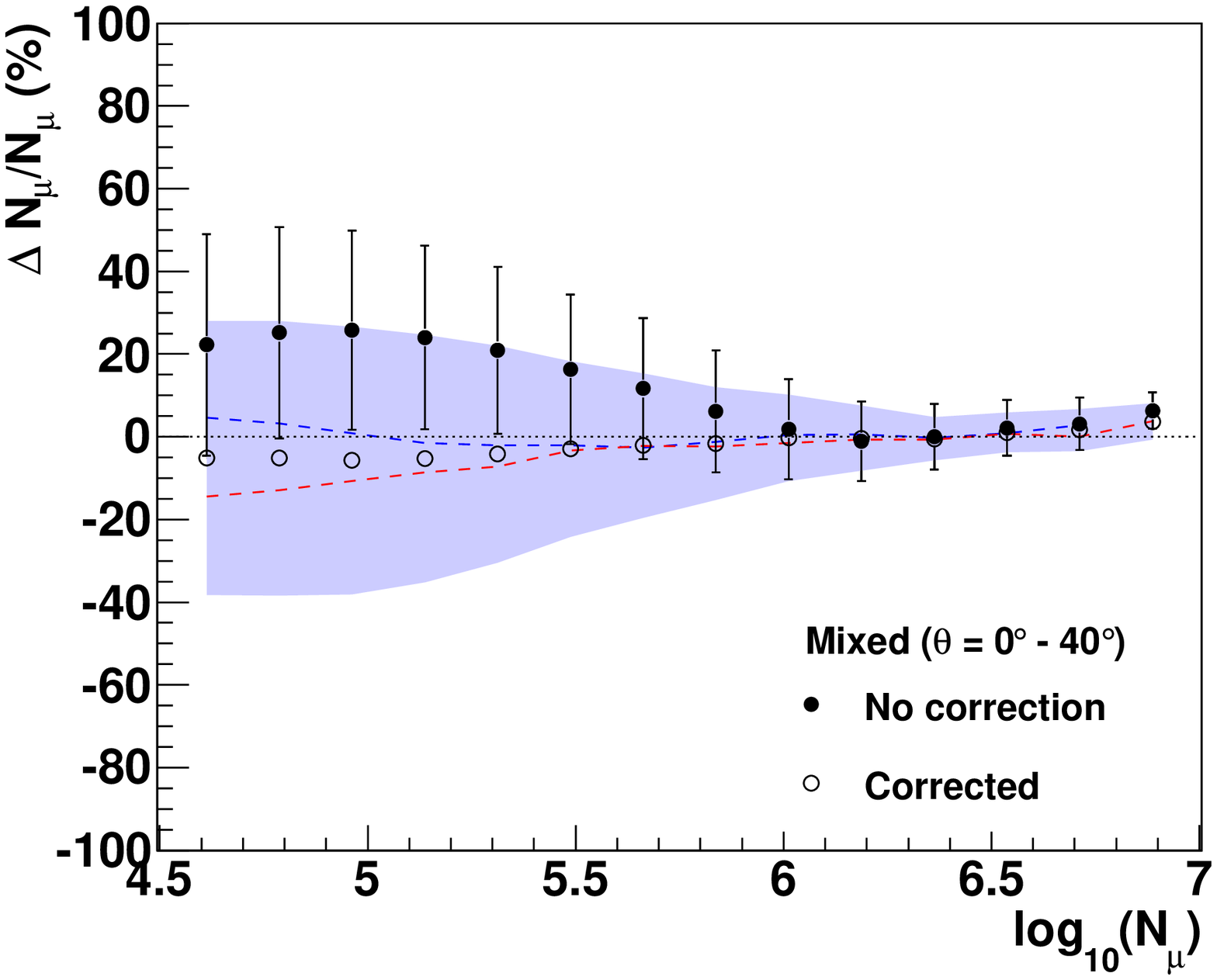}
 \includegraphics[width=75mm, height=60mm]{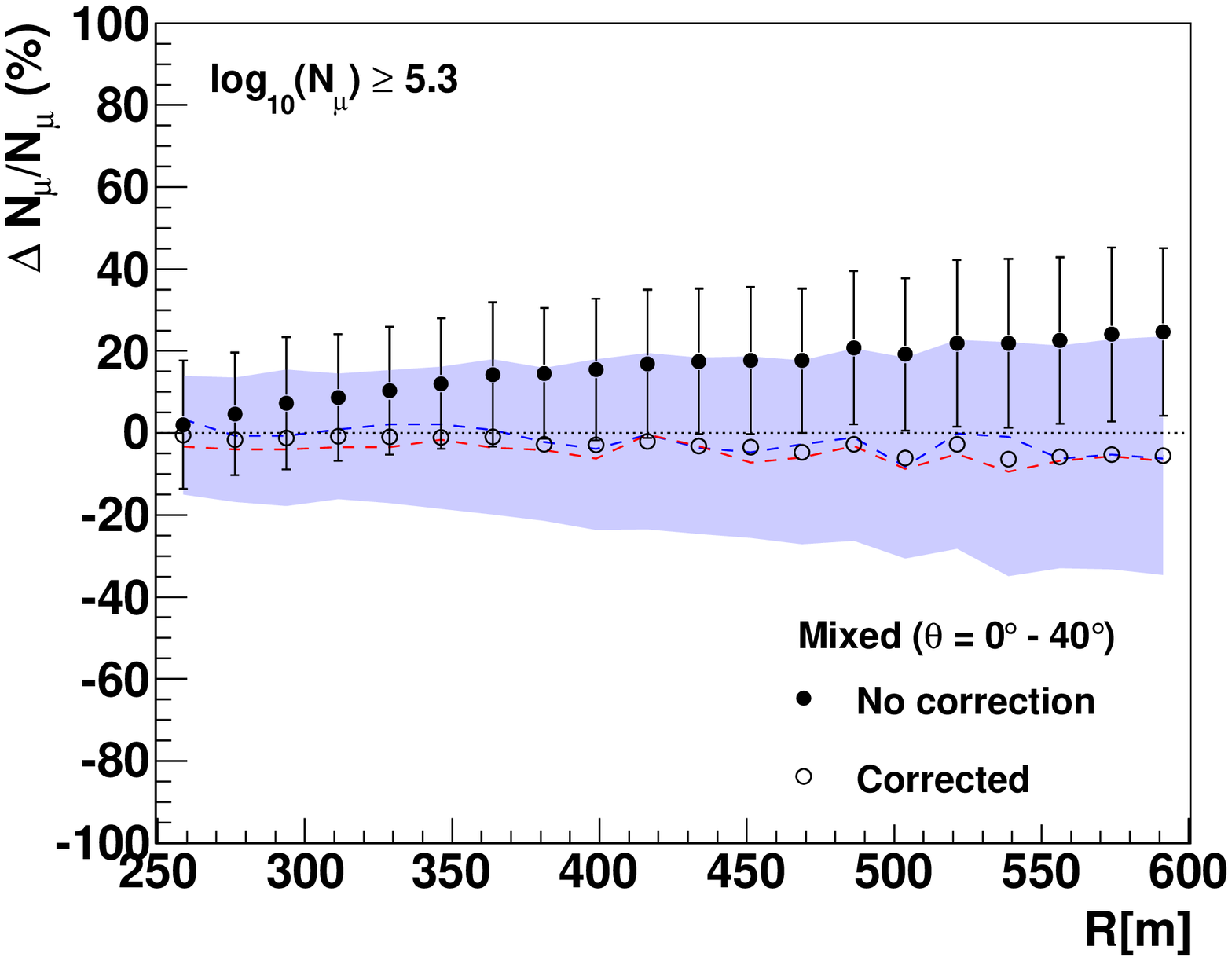}
  \captionsetup{justification=RaggedRight}
 \caption{\small
  Muon systematic uncertainties vs $N_\mu$ (left) and the distance to the
  KASCADE center (right). The situations before  (solid points) and after
  (open circles) applying the muon correction function are compared. Vertical
  lines and grey bands represent the one-sigma errors,
  respectively. Plots were calculated with CORSIKA/QGSJET II for a mixed 
  composition. Dotted lines correspond to pure proton and iron nuclei
  after using the muon correction function (upper and lower dashed lines,
  respectively).}
 \label{fig2}
\end{figure*}

 The same kind of study cannot be applied to the muon size since shielded 
 detectors are only available for KASCADE. In this case, Monte Carlo
 simulations have to be used. The results show that, for $N_{ch} > 10^{4.7}$, the $N_\mu$ systematic 
 uncertainties are $\leq 25 \%$, decreasing for high energies (see figure 
 \ref{fig2}, left), and exhibit a dependence with the core position and the 
 zenith angle (see, for example, figure \ref{fig2}, right). Since the
 behavior of the muon uncertainties are well understood, a muon 
 correction function can be constructed and applied to the muon data.
 This function was built from Monte Carlo simulations for a mixed
 composition assumption and was parameterized as a 
 function of the muon size, the EAS arrival direction and the distance to the 
 KASCADE center. The systematic uncertainties of the corrected muon  number 
 are  $\leq 15\%$ at threshold and quickly decrease for higher muon numbers as 
 seen in figure \ref{fig2}.

 \section{The energy spectrum}
 \subsection{Data sample}

   In order to reduce the influence of systematic uncertainties in the 
 data, a fiducial area located at the center of KASCADE-Grande 
 (as shown in figure \ref{fig1}) was selected and only events with zenith 
 angles $\theta \leq 40^\circ$, which passed the
 KASCADE-Grande reconstruction without failures were considered. 
 Additionally, several experimental cuts were imposed, which results
 in an effective time of observation of 1173 days and an exposure of
 $2.003 \cdot 10^{13} \, \mbox{m}^2 \cdot \mbox{s}\cdot \mbox{sr}$.
 Simulations show that for the above conditions full trigger and 
 reconstruction efficiency at KASCADE-Grande is found above $E \approx 10^{16} \, \mbox{eV}$.
 
  Simulations were employed to study the systematic uncertainties, the 
 performance and reconstruction methods in the experiment as well as
 the influence of the cuts. Both the air shower production and development 
 were realized using CORSIKA \cite{Heck} and the hadronic generators FLUKA
 \cite{Fluka} and QGSJET II \cite{QGSJET}. Events with an homogeneous core 
 distribution and isotropic arrival 
 direction were produced for a $\gamma = -2$ power law spectrum ($E=10^{15} 
 - 3 \cdot 10^{18} \, \mbox{eV}$) and sets were generated for a mixed
 composition of H, He, CO, Si and Fe on similar abundances. A weight function 
 was finally applied to the simulated data to describe a spectrum with $\gamma =
 -3$, which resembles the observed value. 

 \subsection{Strategy}
 
  To reconstruct the energy spectrum, three independent techniques have been 
 applied here: one based solely on the total charged number of particles
 \cite{Kang}, another one on the muon number \cite{arteaga}, and the last
 one on a combination of both observables \cite{bertaina}. 
  
 In the ideal case, when the measurements are accurate enough, when the
 reconstruction procedures work properly, when the Monte 
 Carlo simulations are a faithful description of the EAS and the 
 cosmic ray composition is known one expects the same spectrum from
 all three methods. But since these conditions are not fulfilled, some
 differences are expected among the final experimental results. At this point, 
 the present strategy shows different advantages, it allows (1) to carry out 
 different cross-checks of the reconstruction procedures, the influence of 
 systematic uncertainties and the performance of the detectors, (2) to 
 test the sensitivity of $N_{ch}$ and $N_\mu$ to the elemental composition and 
 (3) to study the validity of hadronic interaction models. In the next
 section, the different reconstruction methods and results will be described.
 It is worth to say that the reconstruction of the energy spectrum has also been
 performed via the density of carged particles at a distance of $500 \, \mbox{m}$
 \cite{Toma}, but the results of this study deserves a more detailed discussion.
 They will be presented in an upcoming paper.

 \subsection{Reconstruction with a single observable}
 
\begin{figure}[!t]
 \centering
 \includegraphics[width=75mm, height=55mm]{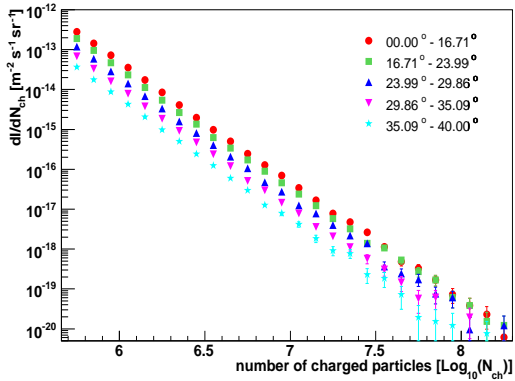}
  \captionsetup{justification=RaggedRight}
 \caption{\small The $N_{ch}$ differential spectra derived from KASCADE-Grande
 measurements.}
 \label{fig3}
\end{figure}

\begin{figure}[!t]
 \centering
 \includegraphics[width=75mm, height=55mm]{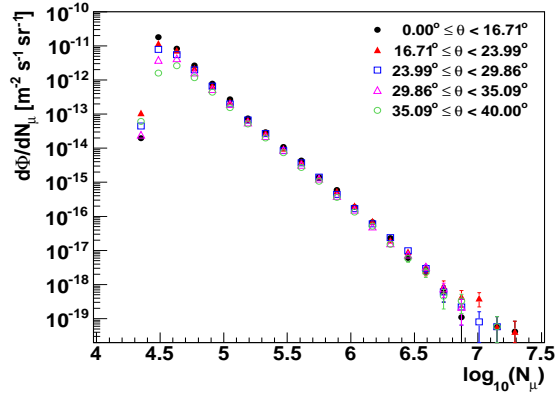}
  \captionsetup{justification=RaggedRight}
 \caption{\small The $N_{\mu}$ differential spectra derived from KASCADE-Grande
 measurements. The muon correction function was applied to the data.}
 \label{fig4}
\end{figure}

 The simplest way to reconstruct the energy spectrum is by using one single 
 parameter, as $N_{ch}$ or $N_\mu$ \cite{Kang, arteaga}. Data is divided in five zenith angle
 intervals with equal acceptance. In each case, the corresponding shower
 size spectrum is extracted (see figures \ref{fig3} and \ref{fig4}) and the 
 constant intensity cut (CIC) method is 
 applied to correct for attenuation effects in the atmosphere. A reference
 angle, $\theta_{ref}$, must be chosen for this task. In this study,
 $\theta_{ref}$ corresponds to the mean of the corresponding zenith angle 
 distribution, which for $N_{ch}$ is $20^{\circ}$ and for $N_\mu$,
 $22^{\circ}$. The difference comes from the particular cuts applied in 
 each case.

 As a next step, from the corrected observable the energy is
 inferred event-by-event by invoking a Monte Carlo calibration 
 function, which is composition dependent. The calibration formula is 
 obtained by fitting with a power law expression, $E = \alpha \cdot N_{ch
 (\mu)}^{\beta}$, the Monte Carlo data points for true energy vs shower size 
 around $\theta_{ref}$ for different mass groups. Then, all data from the 
 distinct zenith angle intervals is combined according to the energy 
 and a single energy spectrum is obtained.
 
 In a final step, the effect of migration of events in the reconstructed 
 energy spectrum is taken into account by applying a response matrix, $R_{ij}$ 
 (also extracted from simulations). If $n^{exp}_{i}$ and $n^{true}_{j}$ are
 the number of events inside the reconstructed and true energy intervals, 
 $\log_{10}{E_{i}}$ and $\log_{10}{E^{true}_{j}}$, respectively, then the 
 response matrix is applied in the following way: $n^{true}_{j} = \sum_{i=1}
 R_{ij} n^{exp}_{i}$, where $\sum_{i=1} R_{ij} = 1$, since it represents the 
 $\log_{10}{E^{true}_{j}}$ probability distribution of an event 
 with reconstructed energy inside the bin $\log_{10}{E_{i}}$. Before 
 unfolding, the response matrix has to be smoothed to avoid the introduction
 of artificial fluctuations due to the limited statistics in Monte Carlo 
 simulations. In figure \ref{fig5}, the unfolded spectra are plotted.

 \subsection{$N_{ch} - N_\mu$ method}

\begin{figure}[!t]
 \centering
 \includegraphics[width=75mm, height=55mm]{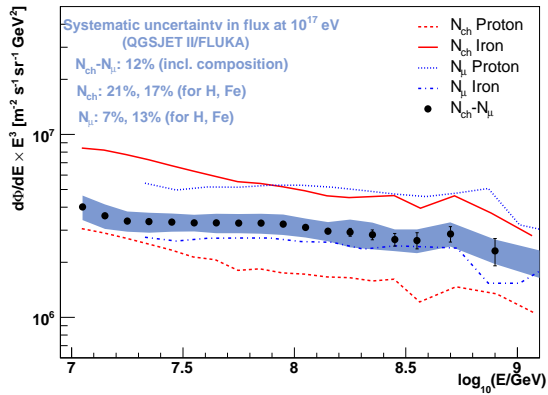}
  \captionsetup{justification=RaggedRight}
 \caption{\small Reconstructed all-particle energy spectrum, multiplied by
 $E^{3}$, as obtained from the KASCADE-Grande data and the different 
 methods described in the present paper. The grey band represents the
 preliminary systematic uncertainty for the spectrum reconstructed through the 
 $N_{ch}-N_{\mu}$ method.}
 \label{fig5}
\end{figure}

  This path to reconstruct the energy spectrum exploits the information from
  both the $N_{ch}$ and $N_\mu$ observables \cite{bertaina}. The energy of an EAS is
  derived through a Monte Carlo expression, which involves the values of 
  the total number of charged particles and the muon size in order to reduce
  the composition dependence of the energy assignment for the EAS. The formula 
  has the form
  \begin{eqnarray}\label{eq1}
   \log_{10}(E/GeV) &=&  [a_p + (a_{Fe} - a_p)\cdot k] \cdot \log_{10}(N_{ch}) \nonumber  \\
                    &+&  [b_p + (b_{Fe} - b_p)\cdot k]
  \end{eqnarray}
  where $k = k(N_{ch}, N_\mu)$ is a mass sensitive parameter defined as
  \begin{equation}\label{eq2}
    k =  \frac{\log_{10}(N_{ch}/N_\mu)      - \log_{10}(N_{ch}/N_\mu)_p}
             {\log_{10}(N_{ch}/N_\mu)_{Fe} - \log_{10}(N_{ch}/N_\mu)_p}
  \end{equation}  
  in such a way that for protons the average value of $k$ is close to zero and 
  approximately one for iron nuclei. The constants ($a$, $b$) and ($c$, $d$)
  are obtained from fits to the scatter plots of $E$ vs $N_{ch}$ and
  $N_{ch}/N_\mu$ vs $N_{ch}$, respectively, and performed in the region of maximum
  efficiency. To take into account the influence of the atmospheric attenuation in 
  the EAS the above formulas are built for every zenith angle interval and
  applied accordingly. As it was the case for the single parameter
  reconstruction, the energy spectrum  obtained by the present method is
  corrected for the effect of the migration of events. The final spectrum
  is shown in figure \ref{fig5}.
  
 \subsection{Flux uncertainties}

  Each method has its own systematic uncertainties, which were carefully 
  estimated when calculating the energy spectrum. In the case of the single 
  parameter reconstruction the sources of systematic uncertainties considered
  for this work were the following ones: the energy calibration relation,  
  the CIC method, the muon correction function, the uncertainty of the 
  spectral index  and the response matrix. For the
  $N_{ch}-N_\mu$ method the contributions to the systematic errors of the
  spectrum come from the atmospheric attenuation, the energy calibration
  function, uncertainties in the primary composition (applying the method to MC
 simulations with different composition assumptions), the spectral slope, 
  and the accuracy of the reconstruction of shower sizes. The total systematic 
  uncertainties of the reconstructed spectra at $10^{17} \, \mbox{eV}$ for the
  different methods are shown in figure \ref{fig5}.

 \section{Discussion}

  As it was pointed out before, the reconstruction of the energy 
 spectrum by using several methods allows to perform several 
 tests and cross-checks. First, by a direct comparison of the
 resulting all-particle energy spectra from the KASCADE-Grande data,
 a good agreement is found inside the respective systematic 
 uncertainties, which means that the experiment and its 
 components are well understood, also that the reconstruction 
 techniques are working as expected and that the hadronic 
 interaction model employed, QGSJET II/FLUKA, is intrinsically 
 consistent. 

 Another interesting observation from figure \ref{fig5}, it is that 
 the muon size at sea level is a very good and composition independent
 energy estimator compared with the total number of charged particles.
 This conclusion is deduced when comparing the difference between the 
 solutions derived from the $N_\mu$ method against the one obtained 
 through the $N_{ch}$ approach alone.

 \begin{figure}[!b]
 \centering
 \includegraphics[width=75mm, height=55mm]{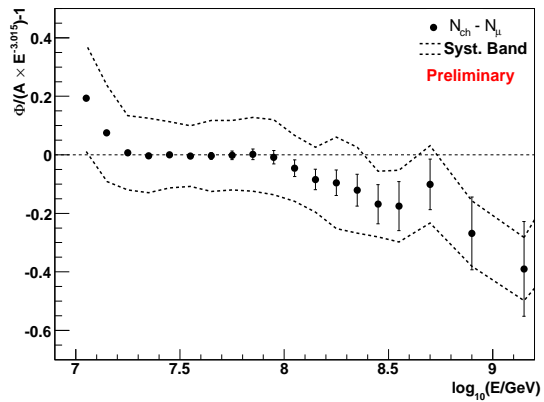}
  \captionsetup{justification=RaggedRight}
 \caption{\small Residual plot for the reconstructed all-particle energy 
 spectrum obtained from KASCADE-Grande through the $N_{ch}-N_{\mu}$ method.
 The systematic error band is also shown (dotted lines).}
 \label{fig6}
\end{figure}

 Some general insights about the composition of cosmic rays at high 
 energies can also be extracted from these graphs. By departing from
 the idea that the true energy spectrum should lie inside the solutions 
 obtained from the $N_\mu$ and the $N_{ch}$ methods, taking into 
 account that the energy spectrum obtained from  $N_\mu$ by assuming 
 pure protons (iron nuclei) is higher (lower) than the one derived
 from $N_{ch}$ and looking at the position of the spectrum from the
 $N_{ch}-N_{\mu}$ method, then it is noticed that the common solution spanned 
 by the different methods favors a relatively heavy composition at high 
 energies inside the QGSJET II/FLUKA framework. 

 To investigate the details of the energy spectrum derived
 from the $N_{ch}-N_{\mu}$ approach, a residual plot was constructed by 
 a direct comparison with a flux proportional to $E^{-3.015}$ (see figure 
 \ref{fig6}). The spectral index of the reference flux was obtained by 
 fitting the middle range of the experimental spectrum, i.e. the 
 interval $E = 10^{16.2} - 10^{17} \, \mbox{eV}$. Two features shows up
 at figure \ref{fig6}, one is a concavity above $10^{16} \, \mbox{eV}$
 and another one is a small break at $\approx 10^{17} \, \mbox{eV}$.
 Both structures are found to be statistically significant. 
 For example, a fit with a power law spectrum
 above $10^{17} \, \mbox{eV}$ gives a spectral index of $\gamma = -3.24
 \pm 0.08$. A preliminary statistical analysis with the $F$-test shows that 
 the significance associated with this change of the spectral index 
 is at a level of $\approx 99.8\%$. 

\begin{figure*}[!t]
 \centering
 \includegraphics[width=150mm]{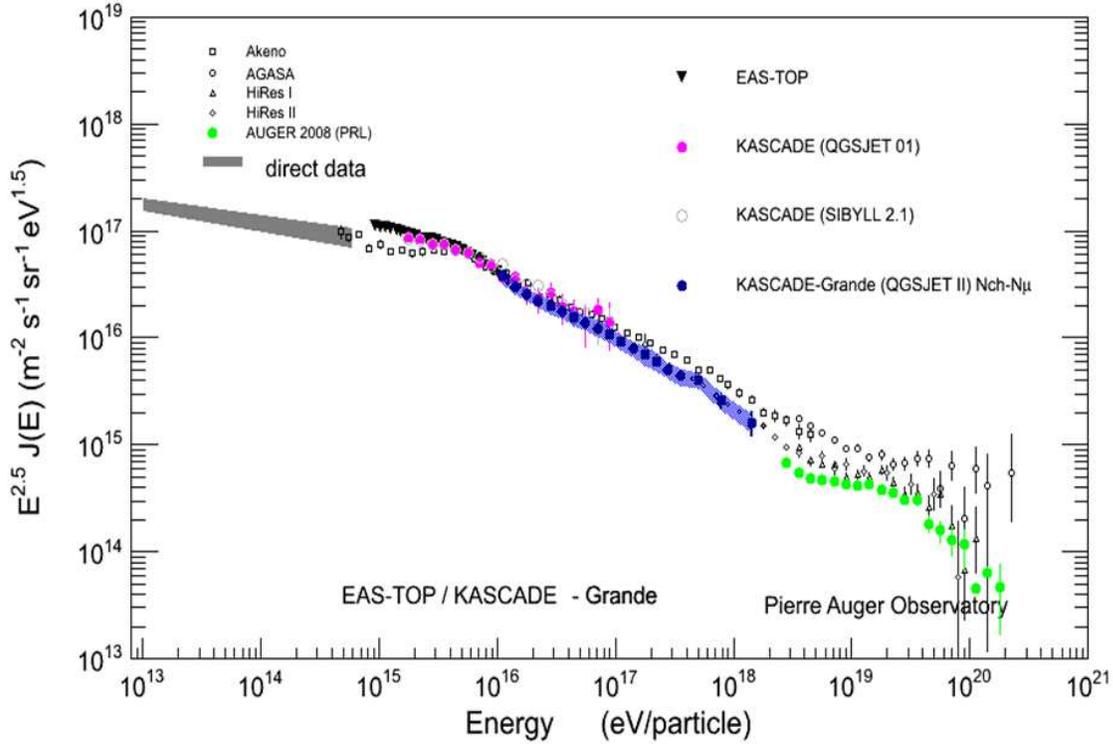}
  \captionsetup{justification=RaggedRight}
 \caption{\small A comparison of the all-particle energy spectrum
  reconstructed from the KASCADE-Grande data through the $N_{ch}-N_{\mu}$
  method with the spectra derived by other experiments.}
 \label{fig7}
\end{figure*}

 In figure \ref{fig7}, the energy spectrum of cosmic rays obtained
 from the KASCADE-Grande measurements is compared with the spectra obtained 
 by other cosmic ray instruments. In general, a good agreement is 
 observed at low and high energies with the KASCADE-Grande data.

 By now the analysis of the influence of the hadronic interaction models is restricted 
 by the statistics of the Monte Carlo data sets generated with alternative hadronic models. 
 However, some preliminary studies applying the reconstruction $N_{ch}$ method \cite{Kang2}
 and the two-observables approach have been already performed with EPOS v1.99 \cite{epos}.
 Those analysis show that the differences between the reconstructed spectra using 
 EPOS v1.99 and QGSJET II are of the order of $10 - 15 \%$. The same preliminary tests done
 with EPOS v1.99 also suggest that the observed stuctures at the energy spectrum reconstructed
 with QGSJET II are not an artifact of the hadronic interaction model.

 \section{Conclusions}
 
  The all-particle energy spectrum of cosmic rays was reconstructed from the 
  KASCADE-Grande data using three different techniques and the hadronic
 interaction models QGSJET II and FLUKA. The resulting energy spectrum 
 shows statistical significant features that can be identified with a 
 concavity and a small break at $\approx 10^{16}$ and $10^{17} \, \mbox{eV}$,
 respectively. 
 The nature of this structures are under discussion. Information about
 the composition in this energy interval will provide valuable clues
 to solve the mistery. These composition studies are now right underway
 at KASCADE-Grande.
 \vspace{1pc}

\begin{acknowledgments}
 {\footnotesize
 This study was partly supported by the DAAD-Proalmex program (2009-2010). 
 J.C. Arteaga acknowledges the partial support from
 CONACyT and the Coordinaci\'on de la Investigaci\'on Cient\'\i fica de
 la Universidad Michoacana.}
\end{acknowledgments}

\end{document}